\documentclass[pra,twocolumn,groupedaddress,showpacs,byrevtex]{revtex4}
\usepackage{graphics,amssymb,amsmath}

\newcommand{\sect}[1]{Sec.~\ref{#1}}
\newcommand{\fig}[1]{Fig.~\ref{#1}}
\newcommand{\tab}[1]{Table.~\ref{#1}}
\newcommand{\eqn}[1]{Eq.~(\ref{#1})}
\newcommand{\eqns}[2]{Eqs.~(\ref{#1}) and (\ref{#2})}
\newcommand{\ket}[1]{| #1 \rangle}
\newcommand{\ketdown}{|\downarrow\rangle}
\newcommand{\ketup}{|\uparrow\rangle}

\newcommand{\be}{\begin{eqnarray}}
\newcommand{\ee}{\end{eqnarray}}

\begin{document}

\title{Implementing the quantum random walk}

\author{B.~C.~Travaglione}
 \email{btrav@physics.uq.edu.au}
\author{G.~J.~Milburn}
 \affiliation{
  Centre for Quantum Computer Technology, University of Queensland,
  St. Lucia, Queensland, Australia
  }

\date{September 17, 2001}

\begin{abstract}

Recently, several groups have investigated quantum analogues of random
walk algorithms, both on a line and on a circle. It has been found that the
quantum versions have markedly different features to the classical versions.
Namely, the variance on the line, and the mixing time on the circle increase
quadratically faster in the quantum versions as compared to the classical
versions. Here, we propose a scheme to implement the quantum random walk on a
line and on a circle in an ion trap quantum computer. 
With current ion trap technology, the number of steps that could be
experimentally implemented will be relatively small.
However, we show how the enhanced features of these walks could be observed 
experimentally. 
In the limit of strong decoherence, the quantum random walk
tends to the classical random walk. By measuring the degree to which the walk
remains `quantum', this algorithm could serve as  
an important benchmarking protocol for ion trap quantum computers.

\end{abstract}
\pacs{03.67.Lx, 32.80.Pj}

\maketitle

\section{Introduction}\label{intro}

The idea that a computational device based on the laws of quantum mechanics 
might be more powerful than a computational device based on classical mechanics
has been around for about two decades \cite{Feynman82}. 
The study of computational devices based upon quantum mechanics is known as
quantum computation. For an introduction to the field, see for example Nielsen
and Chuang \cite{Nielsen00}.
Active research in this field has exploded since the discovery by
Shor \cite{Shor94} that a quantum computer could, in theory, factor large semi-primes
exponentially faster than can currently be done on a classical computer. 
Since Shor's algorithm, Grover has devised an algorithm which can, in
principle, search an
unsorted database quadratically faster than any classical
algorithm \cite{Grover97a}.
However, new quantum algorithms which out perform their classical counterparts 
are proving difficult to find.
One path which is being followed to find new quantum algorithms involves
looking at effective classical algorithmic techniques, and trying to adapt them 
to quantum computation.
Classically, the random walk has found applications in many fields including
astronomy, solid state physics, polymer chemistry and biology. For a review of
the theory and applications for random walks, see for example Barber and
Ninham \cite{Barber70}.
The hope is that a quantum version of the random walk might lead to
applications unavailable classically.
Quantum random walks have been investigated by a number of groups 
\cite{Aharonov93,Farhi98,Nayak00,Aharonov00,Childs01,Mackay01}.
In this paper, we propose a scheme to implement the discrete quantum random
walk on a line \cite{Nayak00} and on a circle \cite{Aharonov00}, using an ion trap quantum computer. 
For a review of ion trap quantum computation see Wineland et al.
\cite{Wineland98}.
With current ion trap technologies, it will not be possible to implement a large
number of steps in the walk, however it should be possible to implement enough
steps to experimentally highlight the differences between the classical and
quantum random walks, providing an important proof of principle.

The structure of this paper is as follows. 
In \sect{qvc} we review the simple models of random walks on both a line and a 
circle, highlighting the differences between the classical and quantum versions
in both cases. 
In \sect{iontrap} we discuss how we shall be representing the algorithms in an 
ion trap quantum computer. 
We then discuss the pulses required to evolve the system, first for the walk on
the line, and then for the walk on a circle. 
Finally, in \sect{measurement} we discuss a relatively simple measurement
procedure which can be used to highlight the difference between the classical
and quantum random walks.

\section{Classical versus quantum random walks}\label{qvc}

Classical random walks can take many different forms, starting from the simple
discrete random walk on a line, to random walks on graphs, to continuous time
random walks, such as brownian motion. In this paper we are only considering
discrete time, discrete space, random walks on a line and on a circle.

\subsection{Classical walk on a line}\label{cwalkline}

Imagine a person standing at the origin of a line with a coin in their hand.
They flip the coin, and if it comes up heads, they take a step to the right, if
it is tails, they take a step to the left. They then repeat this procedure,
flipping the coin, and taking a step based on the result. 
The probability, $P_N(d)$, of being in a position $d$ after $N$
steps is
\be
 P_N(d) &=& \frac{1}{2^N} 
   \left( \! \begin{array}{c} N \\ \frac{d+N}{2} \end{array} \! \right).
\ee 
\tab{classtab} contains the probabilities for the first few values of $N$.
\begin{table}
 \begin{tabular}{|c|c|c|c|c|c|c|c|c|c|} \hline 
 \begin{picture}(12,12)\put(-1.5,12){\line(1,-1){15}}
    \put(-1.7,-2.5){$N$}\put(8,5){$d$} \end{picture}
 &-$4 $&-$3 $&-$2 $&-$1 $&$ 0 $&$ 1 $&$ 2 $&$ 3 $&$ 4 $\\ \hline
 $ 0 $& & & & &$ 1 $& & & & \\ \hline
 $ 1 $& & & &$ \frac{1}{2} $&$ 0 $&$ \frac{1}{2} $& & & \\ \hline
 $ 2 $& & &$ \frac{1}{4} $&$ 0 $&$ \frac{1}{2} $&$ 0 $&$ \frac{1}{4} $& & \\ \hline
 $ 3 $& &$ \frac{1}{8} $&$ 0 $&$ \frac{3}{8} $&$ 0 $&$ \frac{3}{8} $&$ 0 $&$
 \frac{1}{8} $& \\ \hline
 $ 4 $&$ \frac{1}{16} $&$ 0 $&$ \frac{4}{16} $&$ 0 $&$ \frac{6}{16} $&$ 0 $&$
 \frac{4}{16} $&$ 0 $&$ \frac{1}{16} $\\ \hline 
 \end{tabular}
 \caption{The probability of being found at position $d$ after $N$ steps of
 the classical random walk on the line.}
 \label{classtab}
\end{table}
The non-zero elements of the distribution are simply terms from Pascal's 
triangle,
divided by the appropriate factor of two. There are two features of this
random walk that we would like to compare to the quantum analogue. Firstly,
the mean of the walk is zero. This is intuitively obvious, we are using a fair
coin, so we are as likely to step left as we are to step right. 
The other property of the distribution that we are interested in is the
standard deviation. It is not hard to calculate that the standard deviation of
this distribution, $\sigma_c$, is given by
\be
 \sigma_c &=& \sqrt{N}.
\ee

\subsection{Quantum walk on a line}\label{qwalkline}

\begin{figure}[hb]
 \centering
 \scalebox{0.37}{\includegraphics{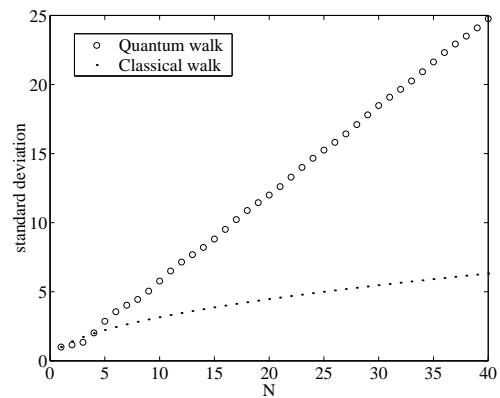}}
 \caption{Standard deviation for both the quantum and classical random walks
          up to $N=40$.}
 \label{std}
\end{figure}

Now let us consider a quantum version of the walk on a line. The first
modification we can make is to replace the coin with a qubit. In this paper,
we shall be representing the two levels of the qubit with the states $\ketdown$
and $\ketup$ rather than $\ket{0}$ and $\ket{1}$. If we start with
the qubit in the down state, and apply a Hadamard operation, we get an equal
superposition of up and down,
\be
 \hat{H}\ketdown = \frac{1}{\sqrt{2}}\ketup + \frac{1}{\sqrt{2}}\ketdown,
  &\quad& \hat{H} = \frac{1}{\sqrt{2}}
     \left(\begin{array}{cc} 1&1\\1&-1\end{array}\right).
\ee  
If we were to measure the qubit, and step left or right depending upon the
result, we would obtain exactly the classical walk described above.
Now, rather than a person holding a coin, suppose we have a particle,
whose motion is confined to one dimension. We can now treat the particle as a
quantum system, and perform the quantum walk as follows. During each
iteration, we apply the Hadamard operation, followed by the operation which
steps right if the qubit is down, and steps left if the qubit is up. That is,
we apply the operator,
\be\label{walkop}
 \hat{U} &=& e^{i\hat{p}\hat{\sigma}_z} \hat{H},
\ee 
where $\hat{p}$ is the momentum operator of the particle confined to one
dimension, and $\hat{\sigma_z}$ is the pauli-$z$ operator acting on the qubit.
Therefore, the state of the system after $N$ steps is
\be
 \ket{\Psi_N} &=& \left(e^{i\hat{p}\hat{\sigma}_z}\right)^N \ket{\Psi_0},
\ee 
where $\ket{\Psi_0}$ is the initial state of the system.
The mean of the distribution produced by this quantum random walk is not
necessarily zero. It is dependent upon the initial state of the qubit.
For example, choosing the initial state of the qubit to be down gives a
non-zero mean after the second step. For the remainder of this paper, we shall
only be considering the distribution created with the initial qubit state
$\frac{1}{\sqrt{2}}\ketdown + \frac{i}{\sqrt{2}}\ketup$ which has a mean of
zero for all values of $N$,
\be
 \ket{\Psi_0} &=& \frac{1}{\sqrt{2}}\ket{0}(\ketdown + i\ketup).
\ee 
\tab{quanttab} contains the probability distribution associated with the first 
few states $\ket{\Psi_N}$.
\begin{table}
 \begin{tabular}{|c|c|c|c|c|c|c|c|c|c|} \hline 
 \begin{picture}(12,12)\put(-1.5,12){\line(1,-1){15}}
    \put(-1.7,-2.5){$N$}\put(8,5){$d$} \end{picture}
 &$ -4 $&$ -3 $&$ -2 $&$ -1 $&$ 0 $&$ 1 $&$ 2 $&$ 3 $&$ 4 $\\ \hline
 $ 0 $& & & & &$ 1 $& & & & \\ \hline
 $ 1 $& & & &$ \frac{1}{2} $&$ 0 $&$ \frac{1}{2} $& & & \\ \hline
 $ 2 $& & &$ \frac{1}{4} $&$ 0 $&$ \frac{1}{2} $&$ 0 $&$ \frac{1}{4} $& & \\ \hline
 $ 3 $& &$ \frac{1}{8} $&$ 0 $&$ \frac{3}{8} $&$ 0 $&$ \frac{3}{8} $&$ 0 $&$
 \frac{1}{8} $& \\ \hline
 $ 4 $&$ \frac{1}{16} $&$ 0 $&$ \frac{6}{16} $&$ 0 $&$ \frac{2}{16} $&$ 0 $&$
 \frac{6}{16} $&$ 0 $&$ \frac{1}{16} $\\ \hline 
 \end{tabular}
 \caption{The probability of being found at position $d$ after $N$ steps of
 the quantum random walk on the line, with the initial qubit state
 $\frac{1}{\sqrt{2}}\ketdown + \frac{i}{\sqrt{2}}\ketup$.}
 \label{quanttab}
\end{table}
The non-zero elements of the distribution are no longer simply terms from
Pascal's triangle which arose in the classical case. The deviations from the
classical distribution are caused by quantum interference effects. 
Now it is no longer simple to calculate the standard deviation of the
distribution. 
However, numerical simulations reveal that the standard
deviation, $\sigma_q$, is almost independent of the initial state of the qubit, 
and is approximately linear in $N$, 
\be
 \sigma_q &\approx& \frac{3}{5} N.
\ee 
The standard deviation is plotted in \fig{std} up to $N=40$ for both the 
classical and quantum walk distributions. 

Clearly, the standard deviation is significantly different for the
quantum and classical random walks on a line. Now let's consider the random
walks which arise when periodic boundary conditions are applied to the random
walks.

\subsection{Classical walk on a circle}\label{cwalkcircle}

In the paper by Aharonov et al. \cite{Aharonov00}, they consider random walks 
on the circle, where the step size is an irrational multiple of $\pi$. 
Here, we shall only be considering the simple distribution which arises when 
the step size is taken to be $\pi/2$. Let us assume that the particle is
initially found, with probability one, at some point on a circle denoted by
$\theta = 0$,
\be\label{startc}
 P_0(\theta = 0) &=& 1. 
\ee
After one step of the algorithm, the classical distribution is given by 
\be\label{odd}
 P_1(\theta) &=& 
   \left\{ \begin{array}{r@{\quad:\quad}l} 0 & \theta = 0,\pi \\ 
      \frac{1}{2} & \theta = \pm \frac{\pi}{2} \end{array} \right. ,
\ee 
and after the second step, 
\be\label{even}
 P_2(\theta) &=& 
   \left\{ \begin{array}{r@{\quad:\quad}l} \frac{1}{2} & \theta = 0,\pi \\ 
      0 & \theta = \pm \frac{\pi}{2} \end{array} \right. .
\ee 
It is not difficult to see that the probability distribution for all
subsequent odd number of steps will be given by \eqn{odd}, and the
distribution for all subsequent even number of steps will be given by
\eqn{even}.

\subsection{Quantum walk on a circle}\label{qwalkcircle}

Let us consider the quantum random walk on a circle. Once again, we start
with the particle at some point on a circle denoted by $\theta = 0$, thus the
initial probability distribution is given by \eqn{startc}. The probability
distributions after one and two steps are also given by \eqns{odd}{even}
respectively, however after the third step, interference effects results in
the distribution
\be
 P_3(\theta = \frac{\pi}{2}) &=& 1. 
\ee
Calculation of the states after subsequent steps reveals that the quantum
random walk around the circle, with a step size of $\pi/2$ is periodic with a
period of eight. The eight probability distributions which arise are given in
\tab{qcirtab}.

\begin{table}
 \begin{tabular}{|c|c|c|c|c|} \hline 
 \begin{picture}(12,12)\put(-1.5,12){\line(1,-1){15}}
    \put(-1.7,-2.5){$N$}\put(8,5){$d$} \end{picture}
 &$ 0 $&$ \frac{\pi}{2} $&$ \pi $&$ -\frac{\pi}{2} $\\ \hline
 $ 0 $&$ 1 $&$ 0 $&$ 0 $&$ 0 $\\ \hline
 $ 1 $&$ 0 $&$ \frac{1}{2} $&$ 0 $&$ \frac{1}{2} $\\ \hline
 $ 2 $&$ \frac{1}{2} $&$ 0 $&$ \frac{1}{2} $&$ 0 $\\ \hline
 $ 3 $&$ 0 $&$ 1 $&$ 0 $&$ 0 $\\ \hline
 $ 4 $&$ 0 $&$ 0 $&$ 1 $&$ 0 $\\ \hline
 $ 5 $&$ 0 $&$ \frac{1}{2} $&$ 0 $&$ \frac{1}{2} $\\ \hline
 $ 6 $&$ \frac{1}{2} $&$ 0 $&$ \frac{1}{2} $&$ 0 $\\ \hline
 $ 7 $&$ 0 $&$ 0 $&$ 0 $&$ 1 $\\ \hline
 \end{tabular}
 \caption{The probability of being found at position $\theta$ after $N$ steps of
 the quantum random walk on the cirle.}
 \label{qcirtab}
\end{table}

\section{Implementing the walks in an ion trap}\label{iontrap}

The analysis thus far has assumed that all operations can be applied without
error and the particle can exist in position eigenstates. 
Now we shall relax these assumptions, and describe how the algorithm
can be implemented in an ion trap.

The ion trap provides a convenient setting for the quantum random walks we
have described, as it contains the required discrete and continuous quantum
variables. For the remainder of this paper we
shall be discussing implementations based on a single $^9\mathrm{Be}^+$ ion, 
confined in a coaxial-resonator radio frequency (RF)-ion trap, as described in
\cite{Monroe96} and references therein.

The preparation involves laser-cooling the ion to the motional and
electronic ground state, $\ket{0}\ketdown$, as described in \cite{Monroe95}.
A sequence of four Raman beam pulses are then applied \cite{Monroe96} to 
create the state
$(\ket{\alpha}\ketdown + \ket{-\alpha}\ketup)/\sqrt{2}$, where $\ket{\alpha}$
denotes the coherent state of the the oscillator,
\be
 \ket{\alpha} &=& \frac{e^{2\alpha_R\alpha_I i}}{\pi^{1/4}} \int dx 
    e^{\sqrt{2}i\alpha_I x}e^{-\frac{1}{2}(x-\sqrt{2}\alpha_R)^2}\ket{x}
\ee
and $\alpha \equiv \alpha_R + i\alpha_I$.

\begin{figure}[ht]
 \centering
 \scalebox{0.43}{\includegraphics{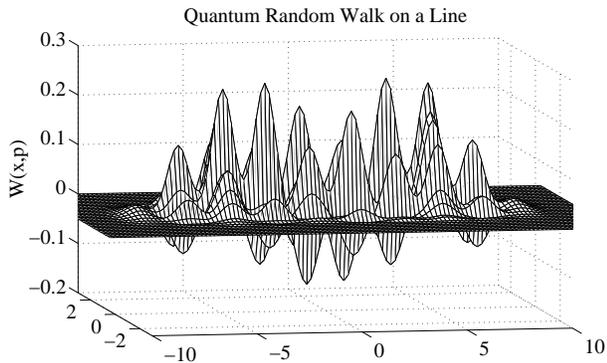}}
 \caption{Wigner function of the particle after five steps of the quantum
 random walk on the line. (The electronic level of the ion has been traced
 over.)}
 \label{wigline}
\end{figure}

The first pulse is a $\pi/2$-pulse which creates an equal superposition of
$\ket{0}\ketdown$ and $\ket{0}\ketup$. A displacement beam is then applied
which excites the motion correlated to the $\ketup$ internal state. The third
pulse is a $\pi$-pulse which exchanges the internal states, and finally the
displacement beam is applied again.
The combined action of the four pulses is to effectively perform 
the operator $\hat{U}$, defined in \eqn{walkop}. The quantum random walk on
the line is accomplished by repeating this sequence of pulses $N$ times. 
\fig{wigline} contains the Wigner function obtained by tracing over the 
internal degree of freedom after five steps of the quantum random walk 
algorithm.

The quantum random walk on the circle can be implemented in an ion trap by
`walking' the particle around a circle in phase space, rather than a circle in
real space. In order to accomplish this task, we need to generate an operator
of the form
\be\label{circop}
 \hat{W} &=& e^{i\pi\hat{a}^\dagger\hat{a}\hat{\sigma}_z/2}\hat{H},
\ee
where $\hat{a}$ and $\hat{a}^\dagger$ correspond to the annihilation and 
creation operators of the harmonic oscillator. This operator can be produced
in an ion trap by applying far-detuned laser pulses to the ion 
\cite{D'Helon96}, followed by a $\pi/2$-pulse.

\section{Measuring the walks}\label{measurement}

Using current ion trap technologies, wave packet dispersion is negligible
\cite{Monroe96}, so the main source of decoherence is related to the internal
levels of the ion. Decoherence of the electronic levels of the ion during the
application of the algorithm has the effect of gradually transforming the 
quantum random walk to the classical random walk. Rather than considering this
to be a negative effect, we can measure the degree to which the ion is acting
as a quantum variable rather than a classical variable, and thereby
effectively measure the level of decoherence in the ion trap.

The scheme that we envisage for measuring the random walk utilises similar 
operators to those employed in the application of the algorithm.
After applying the random walk sequence for some number of steps, the internal
state of the ion is decoupled from the motional state by an appropriate Raman
pulse. An effective operator such as $\exp(i\hat{p}\hat{\sigma}_y)$ is 
applied, before finally measuring the internal state of the ion. 
Thus we are using the internal state of the ion to supply as with information 
about the motional state. 

\begin{figure}[ht]
 \centering
 \scalebox{0.45}{\includegraphics{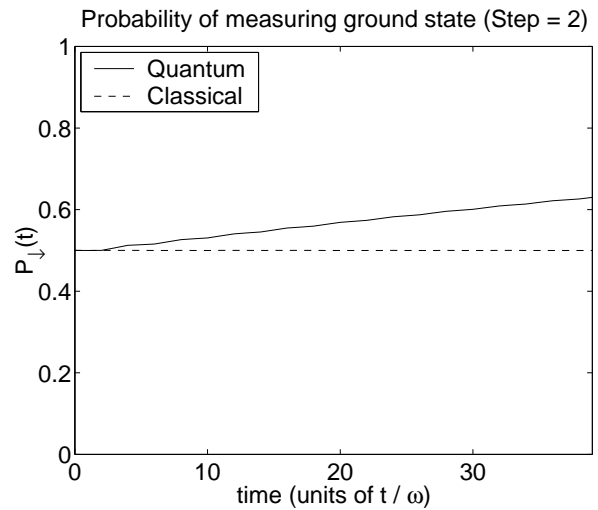}}
 \caption{Probability of measuring the ion in the ground state after applying
 the random walk for a time $t/\omega$, decoupling the internal and motional
 states, and applying the measurement operator $\hat{M}$.}
 \label{qline}
\end{figure}

In the case of the walk on the line, suppose we decouple the internal state from
the motional state by measuring whether the ion is in the state $\ketup$ or
$\ketdown$. We then apply the operator
\be
 \hat{M}^\pm &=& e^{\pm i \hat{p}\hat{\sigma}_y}.
\ee 
The positive Hamiltonian is applied upon obtaining the results $\ketup$,
whilst the negative Hamiltonian is applied otherwise.
Finally, we again measure the
internal state of the ion. If the quantum random walk has experienced no
decoherence, then we measure $\ketdown$ with the probabilities given by the
solid line in \fig{qline}, whereas if the ion suffers complete decoherence we
would expect to measure $\ketdown$ with probability of one half.

\begin{figure}[ht]
 \centering
 \scalebox{0.45}{\includegraphics{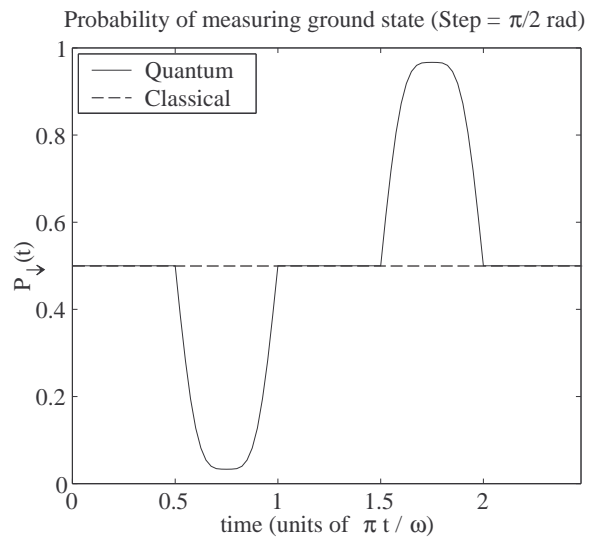}}
 \caption{Probability of measuring the ion in the ground state after applying
 the random walk on a circle for a time $\pi t/\omega$, decoupling the internal
 and motional states, and applying the measurement operator $\hat{D}$.}
 \label{qcircle}
\end{figure}

A similar scheme can be used to measure the level of decoherence in the
quantum random walk on the circle. 
\fig{qcircle} again depicts the probability of measuring the ion in the ground
state after decoupling the internal and motional states, however this time we
then apply the operator
\be
 \hat{D} &=& e^{i \hat{x}\hat{\sigma}_y}.
\ee
In this case, because we have total
destructive interference of certain paths during the walk, the deviation of
the quantum to classical walk is much larger at certain stages of the walk.

\section{Discussion}\label{discuss}

We have described ion trap implementation schemes for quantum random walks,
both on the line and on the circle. We have also suggested a measurement
process which allows the enhanced features of these walks to be experimentally
observed.

At this point, it is unclear whether quantum random walks will have any useful
algorithmic applications. However, we believe that they can provide a
benchmarking protocol for ion trap quantum computers, and perhaps other
implementation schemes which combine continuous and discrete quantum
variables.

\acknowledgments

BCT acknowledges support from the University of Queensland Traveling
Scholarship, and thanks T. Bracken, O. Biham and J. Kempe for useful 
discussions.





\end{document}